\documentclass[twocolumn,amsmath,amssymb,footinbib,superscriptaddress]{revtex4-1}

\usepackage{graphicx}
\usepackage{dcolumn}
\usepackage{bm}
\usepackage{color}
\usepackage{amsmath}
\usepackage[colorlinks=true,citecolor=blue,urlcolor=blue,linkcolor=blue]{hyperref}

\begin{document}

\title{Direct Observation of Sub-picosecond Hole Injection from Lead Halide Perovskite\\by Differential Transient Transmission Spectroscopy}

\author{Kunie Ishioka}
\email{ishioka.kunie@nims.go.jp}
\affiliation{Research Center for Advanced Measurement and Characterization, National Institute for Materials Science, Tsukuba, 305-0047 Japan}

\author{Bobby G. Barker Jr.}
\affiliation{Department of Chemistry and Biochemistry, University of South Carolina, 631 Sumter Street, Columbia, SC, USA}

\author{Masatoshi Yanagida}
\author{Yasuhiro Shirai}
\author{Kenjiro Miyano}
\affiliation{GREEN, National Institute for Materials Science, Namiki 1-1, Tsukuba, 305-00474 Japan}

\date{\today}

\begin{abstract}

Efficient charge separation at the interfaces between the perovskite and with the carrier transport layers is crucial for perovskite solar cells to achieve high power conversion efficiency.  We systematically investigate the hole injection dynamics from MAPbI$_3$ perovskite to three typical hole transport materials (HTMs) PEDOT:PSS, PTAA and NiO$_x$ by means of pump-probe transmission measurements.  We photoexcite only near the MAPbI$_3$/HTM interface or near the back surface, and measure the differential transient transmission between the two excitation configurations to extract the carrier dynamics directly related to the hole injection.  The differential transmission signals directly monitor the hole injections to PTAA and PEDOT:PSS being complete within 1 and 2 ps, respectively, and that to NiO$_x$ exhibiting an additional slow process of 40 ps time scale.  The obtained injection dynamics are discussed in comparison with the device performance of the solar cells containing the same MAPbI$_3$/HTM interfaces.

\end{abstract}

\maketitle


Lead halide perovskite photovoltatic cells have been developing rapidly in the past few years, with their power conversion efficiency (PCE) now exceeding 22\% \cite{NREL}.  The perovskites are direct semiconductors, and their photovoltatics can in principle work as a model \textit{p-i-n} diode \cite{Miyano16ACR}.  The difficulty in the controlled impurity doping in the perovskites can be circumvented by sandwiching the perovskite film between thin layers of electron- and hole-transporting materials (ETM and HTM) in the planar heterojunction structure \cite{Kim16}.  These carrier transporting layers enable efficient and irreversible separation of the electrons and holes photoexcited in the perovskite, and thereby lead to the high PCE of the perovskite solar cells.  Whereas various inorganic and organic materials have been explored as ETM and HTM based on their conduction band minimum (CBM) and valence band maximum (VBM) energies with respect to those of the perovskite, the actual device performance depends very weakly on the energy level offset \cite{Kim16, Belisle16} but can be significantly affected by other factors such as perovskite crystalline quality and interfacial defects \cite{Lim16, Peng16}.

The charge separation dynamics can in principle be monitored directly by means of transient absorption (or transmission) measurements in the visible to teraherz ranges.  The time scale of the charge injection from the perovskite to the HTM layer remain controversial despite the extensive previous studies \cite{Brauer16, Piatkowski15, Ponseca15, Leng16, Corani16, Brauer15, Xing13, Marchioro14, Lim16, Peng16}, however.  The time constant of the hole injection from CH3NH3PbI$_3$ (MAPbI$_3$) to spiro-OMeTAD in the previous reports, for example, scattered widely from $<$80 fs \cite{Brauer16} to 0.7 ps \cite{Piatkowski15, Ponseca15} to 8 ps \cite{Leng16}.  The hole injection to NiO$_x$ was reported to be complete on sub-picosecond time scale \cite{Corani16}, whereas those to poly(triarylamine) (PTAA), poly(3-hexylthiophee-2,5-diyl (P3HT), and poly[2,6-(4,4-bis(2-ethylhexyl)-4H-cyclopenta [2,1-b;3,4-b'] dithiophene)-alt-4,7-(2,1,3-benzothiadiazole) (PCPDTBT) were reported to occur on sub-nanosecond time scales \cite{Brauer15}.  The wide range of the time scale over orders of magnitude appear puzzling at first glance, because the investigated HTMs have relatively similar valence band energies.  The apparent inconsistency among the different studies can be contributed by the different sample preparation methods and different pump and probe light wavelengths employed, as well as by the difficulty in separating hole injection from other carrier dynamics observed in the spectroscopic signals.

\begin{figure}
\includegraphics[width=0.475\textwidth]{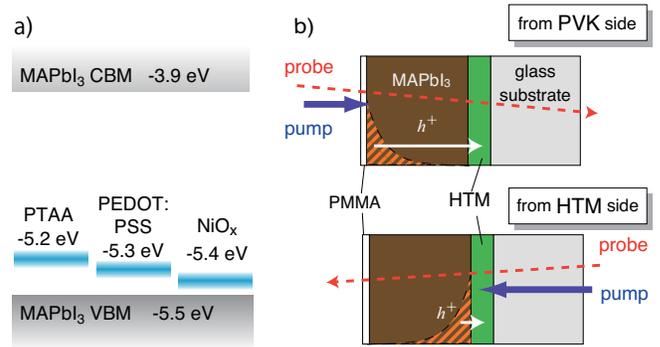}
\caption{\label{Diagram} 
(a) Energy levels of the valence band maxima (VBM) of the HTMs in comparison with the VBM and the conduction band minimum (CBM) of MAPbI$_3$.  (b) Schematics of the two different configurations of the pump-probe transmission measurements of the MAPbI$_3$/HTM sample.  Photoexcited regions inside the MAPbI$_3$ films are designated by hatched area.
}
\end{figure}

In this Letter we propose a simple pump-probe method to extract the carrier injection dynamics at the interface of the perovskite and the carrier transport layer.  We measure transient transmission using excitation light whose optical absorption length is shorter than the perovskite film thickness.  We take the differential transient transmission responses between excitations on the two sides of the samples, and thus directly monitors the interfacial carrier injection dynamics.  We apply this method to the systematic investigation of the hole injection dynamics to the typical organic and inorganic HTMs used in the planar solar cells, PTAA, poly(3,4-ethylenedioxythiophene): poly(styrenesulfonate) (PEDOT:PSS) and NiO$_x$, whose valence band maxima lie at slightly different energies [Fig.~\ref{Diagram}a].  We find the time constant of the hole injection and the quantity of the injected holes depend crucially on the choice of the HTM, and discuss the observed hole injection dynamics in comparison with the device performance of the solar cells containing the same interfaces.


Samples for spectroscopic measurements consist of a glass substrate, a thin layer of HTM prepared by either spin-coating (PTAA and PEDOT:PSS) or sputtering (NiO$_x$) \cite{Yanagida17,Islam17}, and a 250-nm thick crystalline MAPbI$_3$ film \cite{Tripathi15, Khadka17}, as schematically shown in Fig.~\ref{Diagram}b.  For comparison, we also prepare the sample without the HTM layer.  Pump-probe transmission measurements are performed using 400-nm pump and 720-nm probe light pulses with 150-fs durations.  The pump light has an absorption length considerably shorter than the perovskite film thickness  ($\alpha^{-1}$=42 nm \cite{Xie15}) and thus provides an inhomogeneous photoexcitation only near the HTM interface or near the back surface.  The probe light has an absorption length exceeding the film thickness ($\alpha^{-1}$=820 nm \cite{Xie15}) and monitors the carrier dynamics over the whole film thickness.  By taking the difference in the transient transmissions $\Delta T$ measured in two different configurations schematically shown in Fig.~\ref{Diagram}b, we can separate the carrier dynamics directly related to the MAPbI$_3$/HTM interface from other possible effects brought by the HTM.  

\begin{figure}
\includegraphics[width=0.475\textwidth]{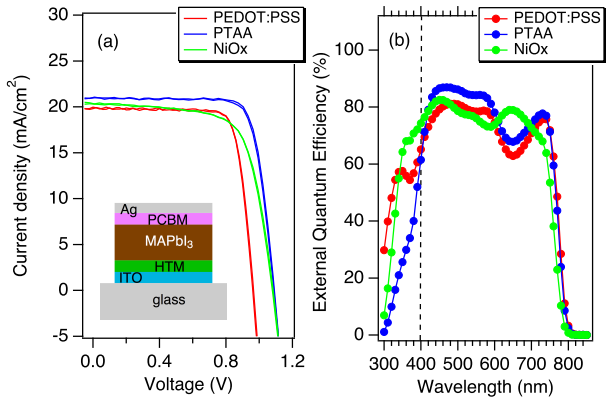}
\caption{\label{JV} Current-voltage curves (a) and external quantum efficiencies (b) for the solar cells with three different HTMs.  Inset in (a) schematically illustrates the solar cell structure.
}
\end{figure}

\begin{table}
\caption{\label{PCE} Device parameters for the the solar cells with three different HTMs obtained from the J-V curves in Fig.~\ref{JV}a; short circuit current ($J_{sc}$), open circuit voltage ($V_{oc}$), fill factor (FF), power conversion efficiency (PCE), series resistance ($R_s$) and shunt resistance ($R_{sh}$).
}
\begin{ruledtabular}
\begin{tabular}{lcccccc}
HTM&$J_{sc}$&$V_{oc}$&FF&PCE&$R_s$&$R_{sh}$\\
& (mA cm$^{-2}$)&(V)&&(\%)&($\Omega$ cm$^{-2}$)&($\Omega$ cm$^{-2}$)\\
\hline
PTAA&20.99&1.09&0.79&18.0&4.89&$>$10000\\
PEDOT:PSS&19.88&0.96&0.79&15.2&3.37&5701\\
NiO$_x$&20.34&1.08&0.69&15.2&5.99&1337\\
\end{tabular}
\end{ruledtabular}
\end{table}

To discuss the hole injection dynamics in comparison with the photovoltatic device performance, we also fabricate the solar cells containing the same MAPbI$_3$/HTM interfaces and measure their current density-voltage (J-V) curves and external quantum efficiency (EQE) \cite{Miyano16JPCL}, whose results are summarized in Fig.~\ref{JV} and Table~\ref{PCE}. We find that the open circuit voltage $V_{oc}$ is highest for the solar cell with PTAA, slightly lower with NiO$_x$, and lowest with PEDOT:PSS, and that the EQE is slightly higher with NiO$_x$ than with PTAA and PEDOT:PSS.  

\begin{figure}
\includegraphics[width=0.475\textwidth]{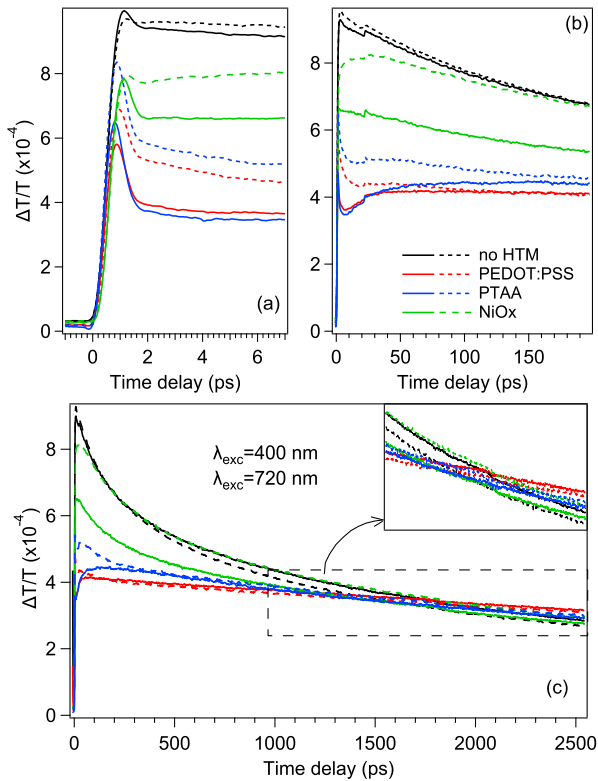}
\caption{\label{DT720} Transient transmission changes of MAPbI$_3$ with different HTMs and without, pumped at 400 nm and probed at 720 nm.  Solid and broken curves denote the transient transmission photoexcited on the HTM and PVK sides, $\Delta T_\textrm{HTM}/T$ and $\Delta T_\textrm{PVK}/T$, as illustrated in Fig.~\ref{Diagram}b.  Pump density is 0.5 $\mu$J/cm2.
}
\end{figure}

Figure~\ref{DT720} compares the transient transmission change $\Delta T/T$ of MAPbI$_3$ with and without HTMs. The transient signals without HTM show no systematic difference between the measurements in the two configurations; they show an instantaneous ($\lesssim$1 ps) rise after photoexcitation, followed by a monotonic decrease that can be fitted to a double exponential decay, $f(t)=A_1\exp(-t/\tau_1)+ A_2\exp(-t/\tau_2)$, with time constants of $\tau_1$=0.16 and $\tau_2$=1.6 ns at pump density of 0.5 $\mu$J/cm$^2$.  The instantaneous rise in $\Delta T$ can be attributed to the photoinduced bleach (PB) of the probe light by the photoexcited carriers.  The faster- and slower-decaying components can be attributed to the non-geminate and Auger recombinations based on their pump density-dependences
.  Our observations are consistent with a previous report on 60-nm thick MAPbI$_3$ film pumped and probed at the same wavelengths \cite{Anand16} as well as with previous time-resolved THz spectroscopic studies \cite{Wehrenfennig14, Johnston16}.  

The presence of a thin layer of HTM modifies the transmission response of the perovskite film drastically.  After the initial instantaneous rise, $\Delta T$ shows almost as rapid decrease [Fig.~\ref{DT720}a].  Moreover, the signals measured from the HTM side ($\Delta T_\textrm{HTM}$) are substantially smaller than that from the PVK side ($\Delta T_\textrm{PVK}$) already at $\sim$1 ps after photoexcitation.  Because the photoexcitation is inhomogeneous along the depth direction [Fig.~\ref{Diagram}b], the smaller $\Delta T_\textrm{HTM}$ can be attributed to the reduction in the PB as a result of the injection of holes photoexcited at the MAPbI$_3$/HTM interface.  The signals from two sides eventually converge, in 100 ps for PEDOT:PSS, 250 ps for PTAA and $\sim$2 ns for NiO$_x$ [Fig.~\ref{DT720}bc].  We attribute the convergence to the holes photoexcited on the PVK side being injected to the HTM interface after diffusing across the perovskite film, since the time scale roughly fits the expectation from theoretical simulations
.  On the even longer time scale, the signals from both sides decay bi-exponentially [Fig.~\ref{DT720}c], like we have already seen for the perovskite without HTM.  The presence of the HTM significantly suppresses the amplitude of the fast component $A_1$ and slow down the decay time $\tau_2$ of the slow component, however.  At the end of our experimental time window (time delay of t=2.5 ns) $\Delta T$ with PTAA and PEDOT:PSS becomes slightly larger than that without HTM, as shown in the insert in Fig.~\ref{DT720}c, reflecting the larger number of electrons surviving the recombination as a result of the deprivation of holes. 


\begin{figure}
\includegraphics[width=0.475\textwidth]{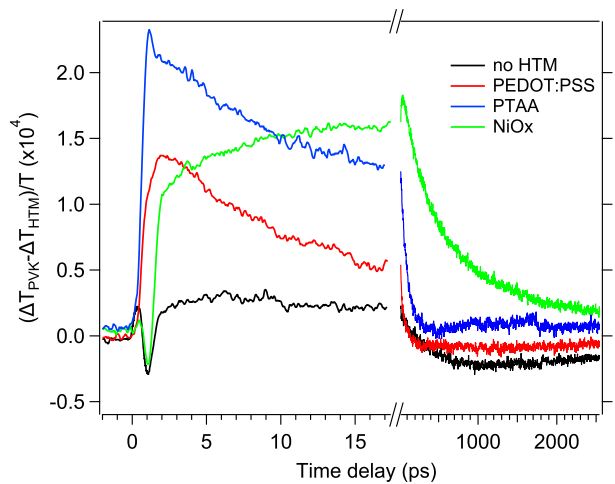}
\caption{\label{Tdiff} Differential transient transmission $\Delta T_\textrm{diff}= \Delta T_\textrm{PVK}-\Delta T_\textrm{HTM}$ of MAPbI$_3$ with different HTMs and without.
}
\end{figure}

Pump-induced transmission changes $\Delta T$ can in principle arise from a variety of ultrafast phenomena including energy-relaxation, diffusion and recombination of photoexcited electrons holes in the MAPbI$_3$ film as well as in the HTM layer.   To extract the carrier dynamics directly related with the MAPbI$_3$/HTM interface and cancel out the others we take the differential transient transmission $\Delta T_\textrm{diff}\equiv \Delta T_\textrm{PVK}-\Delta T_\textrm{HTM}$, as plotted in Fig.~\ref{Tdiff}.  At early time delays when the diffusion can be neglected, this quantity can be regarded as a quantitative measure for the number of holes injected into the HTM.  
We see that $\Delta T_\textrm{diff}$ is very small for the perovskite without HTM, which guarantees the quantitative reproducibility of our measurements; the small spike/dip at t$\sim$0 that may be associated with the defect trapping at the MAPbI$_3$ surface or MAPbI$_3$/glass interface.  In the presence of the PTAA layer, $\Delta$Tdiff rises instantaneously after photoexcitation and reaches a maximum at $t$=1 ps, providing the direct evidence for the hole injection to be complete on sub-picosecond time scale.  $\Delta T_\textrm{diff}$ with PEDOT:PSS exhibits a similar time-evolution, except that it reaches a smaller maximum at a slightly later time ($t$=2 ps) after an apparent two-step rise, though the contribution of the second step is small.  The injection times observed here are comparable with the previous report on the sub-picosecond injection at MAPbI$_3$/spiro-OMeTAD \cite{Piatkowski15, Brauer16}, and orders-of-magnitude faster than the nanosecond hole injection reported in the previous study for the MAPbI$_3$/PTAA photoexcited at a longer wavelength \cite{Brauer15}.  For MAPbI$_3$ with NiO$_x$, the first step of the rise is almost as fast as those with PTAA and PEDOT:PSS, but the second step is considerably slower and contributes larger in proportion; $\Delta T_\textrm{diff}$ reaches a maximum at much later time ($t$=44 ps) as a result.

\begin{figure}
\includegraphics[width=0.475\textwidth]{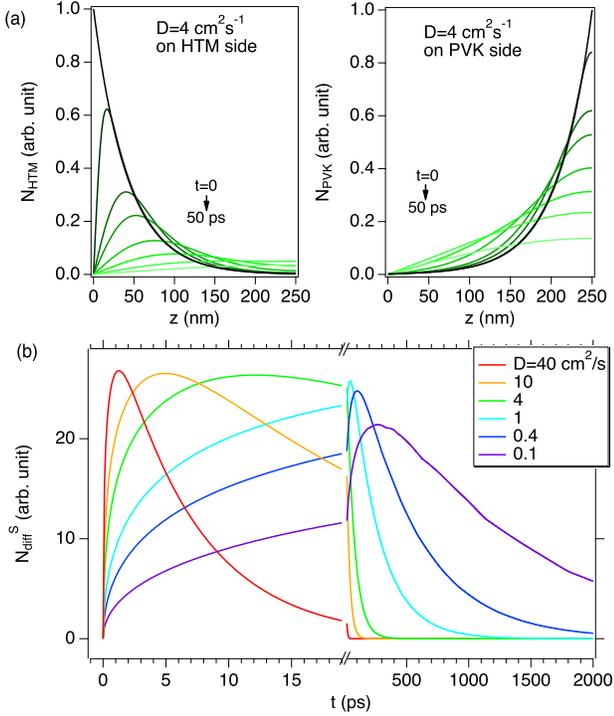}
\caption{\label{DiffInj} (a) Calculated hole distributions $N_\textrm{PVK}$ and $N_\textrm{HTM}$ as a function of distance $z$ from the perovskite/HTM interface for excitation on the PVK and HTM sides at different times $t$.  Diffusion constant $D$=4 cm$^2$/s and recombination time $\tau$=1.6 ns are used.  (c) Calculated differential number of holes $N_\textrm{diff}^S$ as a function of time after photoexcitation for different values of $D$.
}
\end{figure}

The $\Delta T_\textrm{diff}$ signals decay eventually as the holes photoexcited on the PVK side diffuse to the MAPbI$_3$/HTM interface.  The time scale of the decay depends critically on the HTM, however, suggesting that the MAPbI$_3$ films fabricated on top of the different HMT layers have different hole diffusion constant $D$.  Indeed, the previous studies reported different values of $D$, ranging from 0.01 to 4 cm$^2$/s, for different MAPbI$_3$ crystalline qualities \cite{Leng16, Xing13, Guo15, Staub16, Dong15}.  To quantitatively estimate the hole diffusion for the different MAPbI$_3$/HTM samples we perform numerical simulations using different values of $D$
.  Figure~\ref{DiffInj}a shows examples of the hole distributions $N_\textrm{PVK}(z,t)$ and $N_\textrm{HTM}(z,t)$ for excitation on the PVK and HTM sides in the case of $D$=4 cm$^2$/s.  We then obtain the differential hole population $N_\textrm{diff}(t)=N^S_\textrm{PVK}(t)-N^S_\textrm{HTM}(t)$, shown for different values of $D$ in Fig.~\ref{DiffInj}b, where $N_i^S(t)=\int_0^d N_i(z,t)dz$ with $i$= PVK or HTM denotes the total number of holes within the perovskite film at $t$.  We find that $N_\textrm{diff}(t)$ decays on 100 ps time scale, in rough agreement with the decay times of the experimental $\Delta T_\textrm{diff}$ signals for MAPbI$_3$ with PTAA and PEDOT:PSS, when calculated with the largest diffusion constant reported, $D$=4 cm$^2$/s \cite{Dong15}.  The relatively large $D$ confirms the good crystalline quality of these perovskite films.  The \emph{rise} part of the same $N_\textrm{diff}(t)$ trace is much slower than those of $\Delta T_\textrm{diff}$ of the above samples, however.  A possible explanation for the discrepancy is given in terms of the faster hole transport within the photoexcited region near the MAPbI$_3$/HTM interface (hatched area, ``from HTM side" in Fig.~\ref{Diagram}b) than the diffusion in bulk MAPbI$_3$ crystal, possibly driven by the local electronic field in the vicinity of the interface.  A previous photoemission study \cite{Olthof17} reported that the stoichiometry of MAPbI$_3$ within a few nm of the interfaces can be affected by the substrates, and the interface can have a band bending as a result of the dipole formation, in contrast to the commonly assumed flat band condition.  A comparative study on the device performance \cite{Belisle16} also proposed the possibility of ionic accumulation at the interface causing a steep band bending at the interface and a flat band in the rest of perovskite in order to explain the insensitivity of $V_{oc}$ to the VBM of the HTMs.  Such an interfacial band bending can accelerate the hole transport near the interface with PTAA and PEDOT:PSS, and thereby enable the experimentally observed sub-picosecond hole injection to these HTMs.

For MAPbI$_3$ with NiO$_x$, the nanosecond decay of the experimental $\Delta T_\textrm{diff}$ signal can be reproduced by that of $N_\textrm{diff}(t)$ only if we assume smaller diffusion coefficient ($D$=0.4 cm$^2$/s).  It is possible that the MAPbI$_3$ film fabricated on NiO$_x$, a hard inorganic material, contains more defects arising from the lattice mismatch than those on organic PTAA and PEDOT:PSS layers.  The rise part of the experimental $\Delta T_\textrm{diff}$ signal consists of two steps with different time constants; the faster process can be associated with the hole transport accelerated due to the band bending, almost as fast as those at the interface with PTAA and PEDOT:PSS, whereas the slower one can be attributed to indirect injection paths involving defect traps.

Comparison of our spectroscopic results in Fig.~\ref{Tdiff} with the device performance in Table ~\ref{PCE} can give us hint on the choice of the HTM.  The fastest rise and largest maximum amplitude in $\Delta T_\textrm{diff}$ indicate that the holes are injected to PTAA at the fastest rate and the largest quantity, and the solar cell involving PTAA shows the highest $J_{sc}$, $V_{oc}$ and $FF$ despite the large energy offset.  To PEDOT:PSS the holes are injected at slightly slower rate and smaller quantity, and the solar cell exhibits slightly lower $J_{sc}$ and $V_{oc}$.  The hole injection to NiO$_x$ takes an order-of-magnitude longer time than the other two HTMs, possibly because of the interfacial carrier trapping, and the solar cell shows a lower $FF$.  These results suggest a tendency that not only the larger amount of the injected holes but also the faster hole injection in (sub-)picosecond time scale can lead to the better device performance, though we cannot conclude how exactly the carrier dynamics affect these device parameters.


In summary, we have directly monitored the hole injection dynamics at the interfaces of MAPbI$_3$ with three different HTMs by measuring the differential transient transmission signals between the two sides of the sample.  The hole injection to PTAA takes place at the fastest rate and the largest quantity, whereas those to PEDOT:PSS and NiO$_x$ occurs at slightly smaller quantity and at the much slower rate, respectively.  Comparison between the spectroscopic dynamics and the device parameters indicate that both the quantity and the time scale of the hole injection play crucial role in determining the solar cell performance.  The knowledge obtained here will contribute to explore novel HTM materials that enable high photovoltatic performance.





\bibliographystyle{apsrev4-1}
\bibliography{Perovskite}

\end{document}